\documentstyle[amssymb,prb,aps,epsfig]{revtex}
\tighten       
\begin{document}
\draft
\twocolumn[\hsize\textwidth\columnwidth\hsize\csname
@twocolumnfalse\endcsname
\title{Disorder induced phase segregation in $La_{2/3}Ca_{1/3}MnO_{3}$   manganites}
\author{M. Garc\'\i a-Hern\'andez$^1$, A. Mellerg{\aa}rd$^2$, F.J. Mompe\'an$^1$, D. S\'anchez$^1$, A. de Andr\'es$^1$, R.L. McGreevy$^2$ and J.L. Mart\'\i nez$^1$}

\address{$^1$Instituto~de~Ciencia~de~Materiales~de~Madrid (CSIC),Cantoblanco, E-28049~Madrid, Spain}
\address{$^2$NFL Studsvik, Uppsala University, Sweden}
\date{Received \today}
\maketitle

\begin{abstract}
 
Neutron powder diffraction experiments on $La_{2/3}Ca_{1/3}MnO_{3}$    over a broad
 temperature range above and below the metal-insulator transition have been analyzed
 beyond the Rietveld average approach by use of Reverse Monte Carlo modelling.  This 
approach allows the 
calculation of atomic pair distribution functions and spin correlation functions
 constrained to describe the observed Bragg and diffuse nuclear and magnetic scattering.
 The results evidence phase separation within a paramagnetic matrix into ferro and 
antiferromagnetic  domains 
correlated to anistropic lattice distortions in the vicinity of the metal-insulator 
transition.

\end{abstract}




\narrowtext



{\normalsize Rare earth (R) doped manganites with general 
formula $R_{1-x}A_{x}MnO_{3}$, where A is an alkaline earth cation, show an intricate phase diagram with a broad phenomenology
that ranges from 
ferromagnetic (FM) to antiferromagnetic (AF) interactions and from metallic or semiconductor behaviour to charge ordered insulators, depending upon the level of doping and some structural parameters\cite {Coey}.}
{\normalsize However, recent theoretical developments start to draw a unified picture that confers a prominent role to the intrinsic inhomogeinities present in these systems leading to a strong tendency towards phase separation (PS) \cite {Dagotto}. The particular transport and magnetic properties of a given composition would be determined by the specific equilibrium sustained between competing segretated domains: typically ferromagnetic (FM) metallic, antiferromagnetic (AF) orbital or charge ordered, and paramagetic (PM) insulating patches \cite {{Dagotto},{Moreo1},{Yunoki}}.

Large differences observed in the ground state of manganites are related to the existence of 
two different mechanisms leading to the phases mixture: Firstly, electronic phase separation would dominate in the extreme doping regimes and would render nanometer scale clusters. Secondly, equal electronic density phase separation induced by disorder would configure a mixed phase scenario around a first order metal-insulator (MI) transition in the intermediate hole dopping range. Here the level of disorder is the chief variable defining the length scale of each phase. Within the latter context, the MI transition is treated in a percolative framework where the conduction through metallic or insulating phases alternatively dominate the transport below and above $\rm T_{MI}$, respectively \cite{{Moreo},{Mayr}}. Magnetoresistive
properties are also explained as the result of altering the initial equilibrium between phases, the applied magnetic field reinforcing and enlarging the FM and metallic domains at the expense of the insulating ones (either AF or PM)\cite {{Dagotto},{Moreo},{Mayr}}.    

$La_{2/3}Ca_{1/3}MnO_{3}$  (LCMO)  appears to be a good realization of a system where disordered induced PS may take place. It shows a PM to 
FM first order \cite {Rivas} transition around T$_c$= 270 K, about the same temperature where the electronic transport undergoes a MI transition (see Fig. \ref{fig1}). Nonhomogenous states have been inferred in LCMO from experiments supporting polaronic descriptions of the system \cite {{De Teresa},{Ibarra}} but lacking a microscopic 
description of the inhomogenities. Scanning tunneling spectroscopy
has revealed the mixed phase tendencies of LCMO thin films \cite{Fath} but, recently, nanometer PS in thin films has been associated to extrinsic effects, such as the existence of microstrains at the interface with the substrate \cite{Fontcuberta}. Nanometer regions of
charge ordering have been also observed by electron microdiffraction techniques in LCMO \cite{Zuo} and local lattice distortions associated
to polarons have been reported from diffuse x-ray experiments in the PM phase of LCMO \cite {Shimomura}. 

%
This is consistent with Neutron Powder Diffraction (NPD) experiments analized in terms of pair distribution functions that report on local lattice distorsions, compatible with lattice polaron formation as well as with a mixed phase escenario \cite{{Billinge},{Hibble},{Louca}}. These groups, however, do not incorporate in their analysis the diffuse magnetic contribution to the scattering and, consequently, fail to give a microscopic magnetic counterpart associated to the observed lattice distortion. The existence of two such segregated magnetic phases (FM and PM) in LCMO has indeed been reported \cite{Lynn} by inelastic magnetic neutron scattering experiments but, due to the low neutron momentum transfer range explored, these authors cannot provide evidence of the coupling of the magnetic system to the lattice.}
{\normalsize
In  this letter we provide unambiguous experimental evidence of PS at a microscopic
 scale in LCMO and follow the temperature evolution of the interplay of magnetic
 and structural degrees of freedom from 15 K up to 325 K. Our results are compatible with
theoretical predictions \cite {{Dagotto},{Moreo1},{Yunoki},{Moreo},{Mayr}}.     

 We 
have combined Neutron Powder Diffraction techniques 
and Reverse Monte Carlo (RMC) \cite{RMC} modelling of the Bragg (nuclear and magnetic) and diffuse scattering intensities. In RMC modelling, atomic and spin coordinates for a set of ions are arranged non-periodically within a simulation box and varied at random without 
any intervining potential. An optimal configuration for the nuclear
and spin variables is then found whose total (nuclear and magnetic)
calculated powder structure factor, $S(Q)$, fits the observed one in a
least-squares sense. Within this fitted configuration, calculation of
nuclear pair correlations functions, $g(r)$, and spin-spin spatial
$\langle \mu (0).\mu(r´) \rangle$ correlation functions is possible along specified directions and planes in the crystal.


{\normalsize
Polycrystalline $La_{2/3}Ca_{1/3}MnO_{3}$ was prepared by the citrate technique 
\cite{Alonso} and characterized by x-ray diffraction to assess the 
phase purity. The Curie temperature, $\rm T_c$, was determined using SQUID magnetometry to be around 270 K as seen in Fig. \ref{fig1}. NPD data  were measured on the SLAD diffractometer at the NFL Studsvik,
in the temperature range  $15K<T<325K$ and momentum transfer Q range $1<Q(\AA^{-1})<10$.Careful measurements of the empty container,
background and proper substraction of these contributions as well as those from multiple scattering were done. Rietveld refinements of the lattice constants were done for a $Pnma$ cell at all measured temperatures  and the derived atomic coordinates used as starting configuration for the RMC runs at each temperature.}
  
{\normalsize   For the RMC simulations we used a box equivalent to 6x4x6 $Pnma$ ($b$ long axis) single cells for which the La and Ca atoms are treated from the point of view of neutron scattering as a weighted average atom. Classical spins are attached to the positions occupied by magnetic ions whose orientation is also randomly tried. The scattering contribution from this magnetic simulation box is added to the nuclear one in order to make the comparison with the measured structure factor. Only Mn ions carry classical spins whose form factor corresponds to a weighted $\rm Mn^{3+}/Mn^{4+}$ average. This is equivalent to assume equal electronic density for the various phases as stated in the theory \cite {{Moreo},{Mayr}}. For each temperature, all nuclear and magnetic cells are initially dressed with atoms and spins as obtained from Rietveld refinement of the experimental data. Minimum interatomic distances and total magnetic moment constrains are applied. The RMCPOW code \cite{RMC} also provides for the adequate polycrystalline  averaging. 

Figure \ref{fig2}a, shows a representative set of experimental data along
 with the corresponding RMC and Rietveld fittings. Notice that RMC not only shows a better agreement than the Rietveld
refinement but also provides a microscopic decription of the system free from "a prioristic" assumptions, as seen in Figure \ref{fig2}b. 
In order to stress the sensitivity of our
 RMC models to the local magnetic correlations, we have calculated
 the magnetic diffuse scattering for a model in which the magnetic moments
 have been randomly reassigned to different
nuclear sites. This model therefore has the same
 magnetic long-range order (i.e. Bragg peaks), but different
short range order, the resulting diffuse scattering exhibit magnetic intensity 
in the wrong places as shown  Fig. \ref{fig2}c. 

{\normalsize 
We have computed $g(r)_{MnMn}$ for the fitted RMC configurations for Mn-Mn pairs both in the $(a,c)$-plane and along the long axis $b$ (Fig. \ref{fig3}a ). These planes and directions are treated as volumes with adjustable dimensions in the 3-dimensional configuration space. 
Fig. \ref{fig3}b shows the normalized contributions of the in plane $(a,c)$ and out of plane Mn-Mn  to $g(r)_{MnMn}$. It can be seen that nearest neighbours (NN) along different directions exhibit distance distributions centered around the same values (in agreement with Rietveld averages). However, the atom pairs in the $(a,c)$ plane seem to account for the tails of the NN distance contributions, while pairs of atoms along the long axis have a narrower distribution. It must be concluded, therefore, that not only  distortions with respect to the averaged structure exist but these are basically localized within the $(a,c)$-plane. 

Similarly, we have calculated the magnetic moment pair correlation
function, $\langle \mu (0).\mu (r) \rangle$. As expected, the long
distance asymptotic limit of these correlations follow the
macroscopically measured magnetization. That is, above T$_c$ the
system is a paramagnet ( $\langle \mu (0).\mu (r) \rangle$ oscillates
around zero for large $r$), while at low temperatures is a ferromagnet
($\langle \mu (0).\mu (r)\rangle$ $>0$ for large $r$). However, this
scenario evolves at shorter distances through $\rm T_c$ where there is
a strong change in the average character of the short-range magnetic
interactions with the nearest neighbour (NN) and next near neighbours
(NNN) ions. Focusing on the short ranged correlations in
Fig. \ref{fig4}a it is apparent that, on approaching $\rm T_c$ from
above for NN and NNN Mn-Mn pairs, FM correlations build up within the
paramagnetic matrix. Notice, as shown in Fig. \ref{fig2}, that at T =
280 K  small FM Bragg peaks start to develop, implying the existence
of already large FM domains (around 100 \AA) before the system reaches
the metallic regime. On lowering the temperature further, these FM
domains grow to the point in which the basic background is completely
FM below $\rm T_c$. Most interestingly, in addition to these FM
correlations, we also observe that for those Mn-Mn pairs at the larger
distances with respect to the Rietveld average distance $\langle \mu
(0).\mu (r) \rangle$ gets into negative values around $\rm {T_c}$ (See Fig \ref{fig4}b). This suggests the existence of magnetic correlations of an AFM character in this material with typical coherent lengths not larger that about 4-6 \AA. 


Figure \ref{fig1} shows the normalized integrated intensities corresponding to the positive and negative lobes of
the NN spin correlations since it is mostly for these that we have enough pairs (about 50 in the tails) to compute quantities that are 
sound on an statistical basis. Notice that the growth of a negative lobe in 
$\langle \mu (0).\mu (r) \rangle$ does not correlate with the temperature dependence 
three dimensional distortion of the octahedra derived from the 
Rietveld refinement, but it correlates well with 
the calculated 
from the Mn-O pair distribution function of the RMC  configurations (see Fig. \ref{fig1}).
 Furthermore, from the structural insight gained with the RMC, we can establish that,
 on average, the Mn pairs AFM coupled  are located in the $(a,c)$ plane (see Fig. \ref{fig3}b), while those pairs at a distance closer to the average structure (Rietveld value) are along the long axis $b$. As a general trend, it is also observed that those Mn atoms exhibiting AFM interactions are located at larger distances from the intermediate oxygen than its FM counterparts. 

In conclusion, our RMC analysis of the Bragg and diffuse neutron diffraction
intensities draws an inhomogeneous escenario in which three different phases
coexist (FM and AFM domains within a PM matrix) in LCMO near $\rm T_{MI}$.
This microsegregation is probably driven by the 
distortion in the positions of the Mn atoms within the $(a,c)$ plane
that develops around $\rm T_c$ and $\rm T_{MI}$. As a result,
there are Mn atoms whose distance to the NN Mn atom is shorter than that
corresponding to the averaged structure. These atoms present also shorter
Mn-O distances and above T$_c$ hold short range FM interactions within the PM matrix.
As the temperature decreases below $\rm T_c$ these FM clusters increase their coherence
length and the matrix becomes FM. Also, within the $(a,c)$ plane we have found a
significant number of Mn atoms separated from their Mn NN a distance longer than
the averaged Rietveld distance. These pairs (about 15\% of the NN  at 280 K) interact antiferromagnetically around
 $\rm T_c$ and display longer Mn-O distances making double exchange
less efficient as overlap decreases. It is worth mentioning that the AFM domains 
show a phenomenology similar to that reported for the CE-state in $La_{0.5}Ca_{0.5}MnO_{3}$
near the Neel temperature \cite{Radaelli}, making plausible to identify these islands as probable
precursors of the CE-phase.      
Notice also that the onset of the metallic state does not
occur until the AFM correlations considerably weaken, pointing out that the
AFM entities act as strong electron scatterers that add extra restrictions to
the electron transport in the PM phase. It is remarkable that in the 
corresponding Sr doped system,  which
remains metallic above $\rm T_c$, local AFM correlations are far weaker
and $\langle \mu (0).\mu (r) \rangle$ remains
positive \cite {Mellergard}. It is realistic also to think that the
 application of an external field would inhibit the formation of this
 exotic AFM phase resulting in the decrease of the electron scattering
 and the lowering the resistivity.
The picture emerging from our RMC analysis is, therefore, fully consistent with the most recent developments in the theory of manganites that postulate equal density phase separation
induced by disorder\cite{{Dagotto},{Yunoki},{Moreo},{Mayr}}.
The simplified Ising model used by Moreo {\it et al.} contains the basic
 ingredients to account for the observed PS around metal insulator transitions
 predicted, but real systems seem to show an even richer phenomenology 
(three phases are experimentally observed versus the two phases predicted).
 Due to the small lengthscale of the observed FM and AFM domains, the question 
of whether these can be considered as proper thermodynamic phases is open to debate.
 In line with previous works, one could also invoke magnetic polarons as the distorting 
entities but then a reasonable hypothesis has to be added to the theory in order to
 account for the different short range correlations (FM as well as AF). Finally, we
 point out that our results also support the approaches based on more thermodynamic 
grounds that predict the presence of significant AF couplings and phase segregation for
 double exchange systems \cite{Guinea}.

We are grateful to M.T. Casais and M. J. Mart\'\i nez-Lope for preparing the sample.
We acknowledge financial support from the Spanish MCyT (Project MAT99-1045) and the EU - Access to Research Infrastructure
 Action  (HPRI-CT-1999-00061) to enable our access to the NFL-Studsvik.
Critical reading of the manuscript by F. Guinea is greatly acknowledged.

\begin{figure}
        
\caption {a): Temperature variation of the orthorombic distortion in  the RMC local and Rietveld variables,
as defined in the text. b): Temperature dependence of the 
Magnetization (left axis);Normalized integrated intensities (right axis) of the FM lobe, A(+),(open circles) and of the AFM lobe, A(-), (solid circles) of the spin correlations of NN Mn-Mn pairs. 
Inset shows the first derivative of the measured resistivity of LCMO. }   
\label{fig1}
\end{figure}

\begin{figure}
\caption{ a) NPD pattern (circles) for T= 280K along with the corresponding residuals for the conventional Rietveld
refinement(middle profile) and the corresponding for our RMC approach (lower profile).b)
Solid circles are the experimental points, grey areas are  RMC contribution for the magnetic
scattering (diffuse + Bragg),
dashed lines are for the nuclear diffuse scattering and black continuous lines are the sum
of all RMC contributions( Magnetic diffuse, Magnetic Bragg, Nuclear diffuse and Nuclear Bragg).
c)A close up of the results
at 260K pointing out the differences in the magnetic diffuse
scattering corresponding to a  non fitted configuration (thick
line), obtained as explained in the text, along with the corresponding contributions
of the fitted configuration, symbols as in b).}
\label{fig2}
\end{figure}
\begin{figure}
\caption{Upper frame: Distribution function for Mn-Mn pairs, $g_{MnMn}(r)$. Lower frame:
 In ${(a,c)}-$plane (continuous line) and out of plane contributions (dashed line) to $g_{MnMn-total}(r)$ for the
 Mn-Mn pairs. The partial distribution fuction is now normalized
to the Mn-Mn pair distribution function, computed without geometrical 
restrictions, $g_{MnMn-total}(r)$.}
\label{fig3}
\end{figure}
\begin{figure}
\caption{RMC spin correlation functions {\it vs.} distances between Mn-Mn pairs, $r$, for 
all the temperatures explored. Upper frame shows the long distance asymptotic limits, that
coincide with the macroscopic magnetic behaviour. Lower
frame: Detail of the spin correlations for NN and NNN manganese pairs.}  
\label{fig4}
\end{figure}

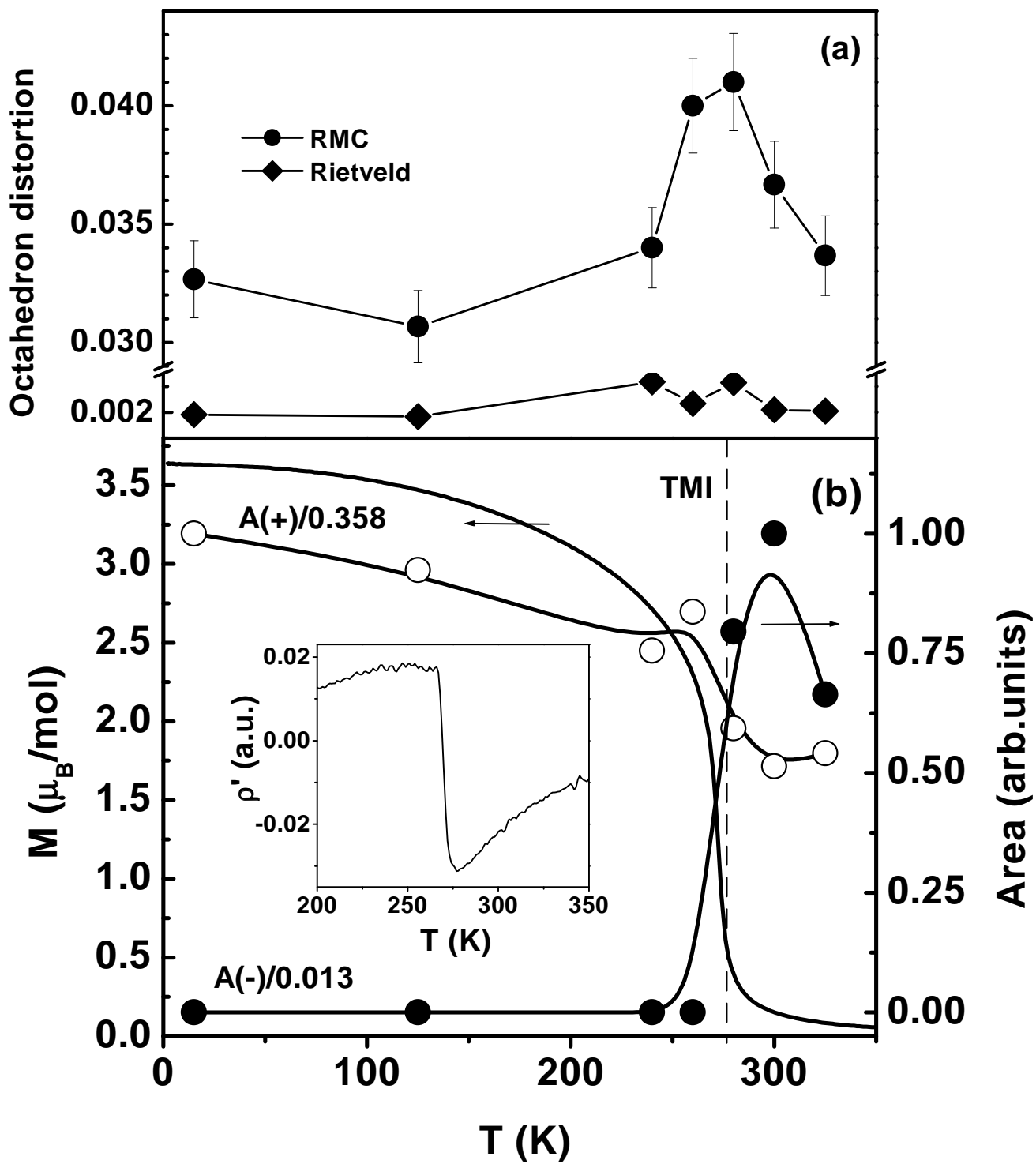
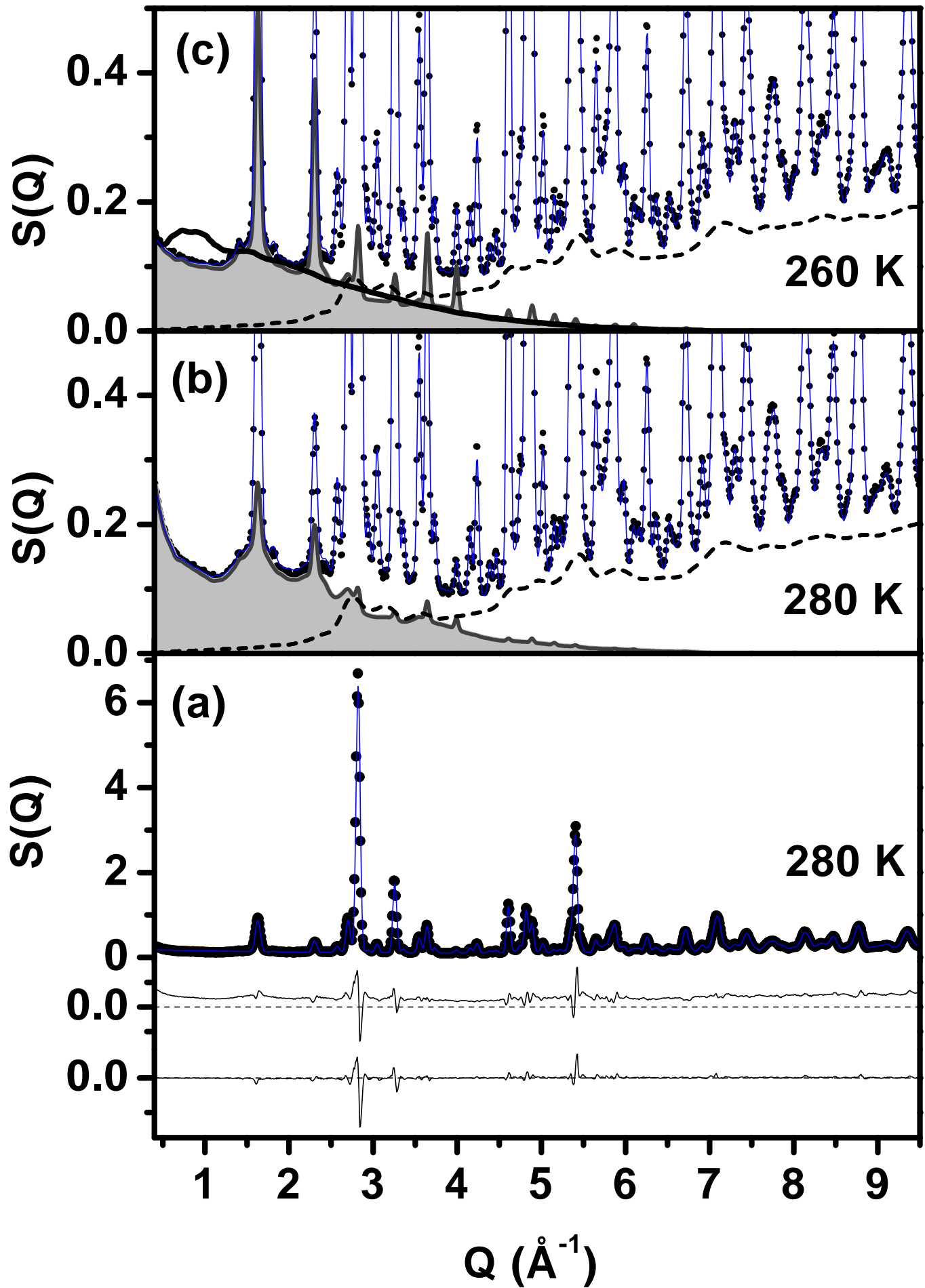
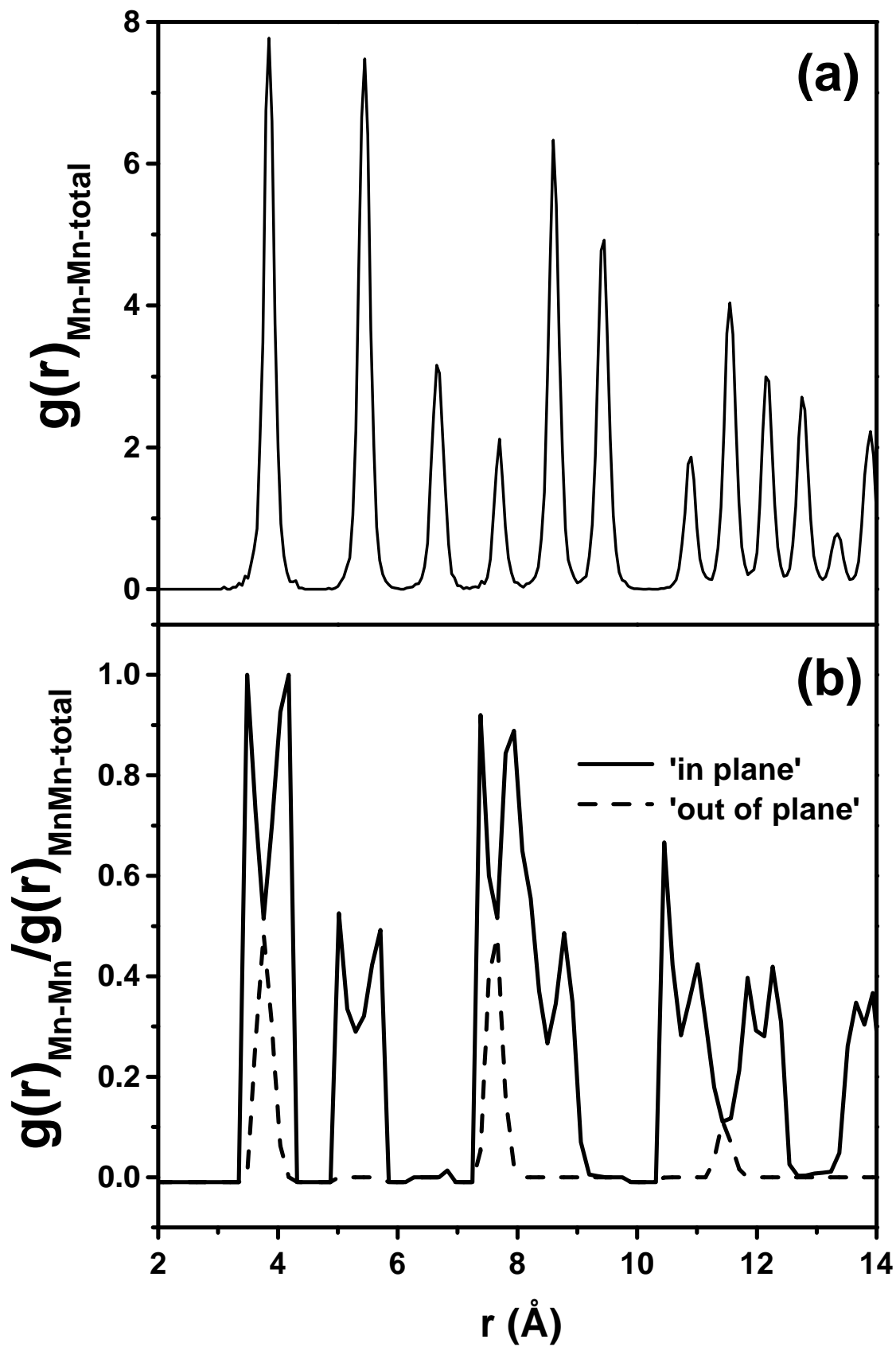
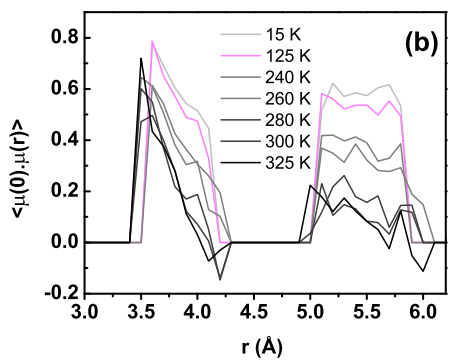
\end{document}